\documentclass{aa}
\usepackage{psfig}
\usepackage{graphicx}
\def\src{SAX J0635+0533}
\def\gsrc{2EG J0635+0521}
\def\deg{\ifmmode^{\circ}\else$^{\circ}$\fi} 

\def\farcs{\hbox{$\,.\!\!^{\prime\prime}$}}
\def\fs{\hbox{$\,.\!\!^{\rm s}$}}
\def\lapp{\ifmmode\stackrel{<}{_{\sim}}\else$\stackrel{<}{_{\sim}}$\fi}
\def\gapp{\ifmmode\stackrel{>}{_{\sim}}\else$\stackrel{>}{_{\sim}}$\fi}

\begin{document}

\title{Radio observations of the 33.8 ms X-ray pulsar \src}

\author{L. Nicastro \inst{1}
\and B.M. Gaensler \inst{2}\thanks{Hubble Fellow}
\and M.A. McLaughlin \inst{3}
}

\institute{
 {Istituto di Fisica Cosmica con Applicazioni all'Informatica, CNR,
 Via U. La Malfa 153, I-90146 Palermo, Italy}
\and 
 {Center for Space Research, Massachusetts Institute of Technology,
  Cambridge, MA 02139, USA}
\and 
{Astronomy Department, Cornell University, Ithaca, NY 14850, USA}
}
\offprints{nicastro@ifcai.pa.cnr.it}
\date{Received 20 June 2000 / Accepted 20 June 2000}
\thesaurus{08(08.09.02 \src; 08.14.1; 08.16.6; 13.18.5; 13.25.5)}
\maketitle
\markboth{L. Nicastro et al.: Radio observations of \src}
         {L. Nicastro et al.: Radio observations of \src}

\begin{abstract}
The hard X-ray source \src\ shows pulsations at a period of 33.8 ms
and is thought to be associated with a bright Be star.
We here report on radio observations towards this source with
the VLA and Arecibo radiotelescopes.
With the VLA we do not detect any radio flux at the Be-star position,
but one 4.7 mJy (at a radio frequency of 1.4 GHz)
source lies just on the border of the $30''$ X-ray error
circle. Periodicity searches at 430 MHz performed with the Arecibo
telescope failed to detect any pulsation.
We argue that the 4.7~mJy source is not the radio counterpart of \src.
Most likely \src\ is a rather ``non-standard''
Be-star/neutron-star X-ray binary pulsar, probably with a relatively short
orbital period. It could still be a radio emitting pulsar but with an
emission beam not pointing toward the Earth.

\end{abstract}

\section{Introduction}

A BeppoSAX observation of the field centered on the unidentified
EGRET source \gsrc\ led to the discovery
of a hard X-ray source, proposed to be associated with a bright
($R\simeq12$) Be-star located within the $1'$ X-ray error circle
(\cite{kaaret}).
Optical spectra showed it is at a distance of 2.5--5 kpc, with an $E(B-V)$
derived neutral hydrogen column density of $6\times 10^{21}$ cm$^{-2}$
(in agreement with the Galactic 21-cm column density toward the source).
The X-ray photon spectrum was fitted with a power law of spectral
index $-1.5$ and column density $N_{\rm H}\simeq 2\times 10^{22}$ cm$^{-2}$.
The derived 2--10 keV X-ray luminosity is $L_X\simeq 23\times 10^{33}
(d_{\rm kpc}/4)^2$ erg s$^{-1}$.
The extra absorption, with respect to the value obtained through
the optical data, suggests that local extra gas must be present.
This interpretation agrees with the Be-star association and led Kaaret et al.
(1999) to the conclusion that the system is a (probably gamma-ray emitting)
high mass X-ray binary.

A reduced X-ray error circle of $30''$, including the Be-star,
resulted from the re-analysis of the BeppoSAX data by Cusumano et al. (2000),
but the most relevant result was the
discovery of coherent pulsed emission at a period of 33.8 ms with a 
significance level of $4\sigma$.
Unfortunately the available X-ray data did not allow them to derive
the system orbital parameters. However, the derived X-ray luminosity
($7.7\times 10^{34} (d_{\rm kpc}/4)^2$ erg s$^{-1}$ in 0.1--40 keV)
and magnetic field strength ($\lapp 10^9$ G), derived from assuming the
pulsar is powered by accretion, are not typical of Be/X-ray binaries
(which have higher luminosities and magnetic fields).

In an accretion scenario, any radio emission is
expected to be quenched within the high density environment.
In a non-accretion (rotation-powered) scenario, high-energy
emission can be produced either by shock acceleration or by
a magnetospheric process. In this case the source could also
show variable and/or periodic radio emission.
Detection of radio pulses
with the same period of the X-ray emission would lead to an
unambiguous interpretation in this sense.

Here we present the results of observations made with the VLA at 1.4 and 4.8
GHz and pulse searches performed using the Arecibo radiotelescope at 430 MHz.

\section{Observations ad data reduction}
The \src\ field was observed on 2000 April 25 with the Very Large Array in
its C configuration at 1.465 and 4.86 GHz; bandwidths were 22 and 100
MHz, respectively.
A gain calibrator source (4C~10.20) was observed at each frequency switch (six
in total) and 3C286 (flux densities of 6.8 and 3.3 Jy at 1.4 and 4.8 GHz
respectively) was used as the primary calibrator.
The total net observing time was $\sim 80$ and
$\sim 110$ minutes at 1.4 and 4.8 GHz respectively.
All four Stokes parameters were recorded.
One of the 27 antennas could not be used because of technical reasons.
Data were processed in the {\tt MIRIAD} package.
After appropriate editing and calibrating, images were formed using
uniform weighting and deconvolved using {\tt CLEAN}.
The synthesized beam is $11''\times 11''$ (1.4 GHz) and
$3\farcs5\times 3\farcs5$ (4.8 GHz).
The resulting images have sensitivities of 120 and 20 $\mu$Jy
at 1.4 and 4.8 GHz, respectively, yielding $5\sigma$ detection limits of
0.6 and 0.1 mJy. In the polarisation images (Stokes Q and U) these limits
are of 0.45 (1.4 GHz) and 0.12 mJy (4.8 GHz).

\begin{figure}[tb]
\centerline{\psfig{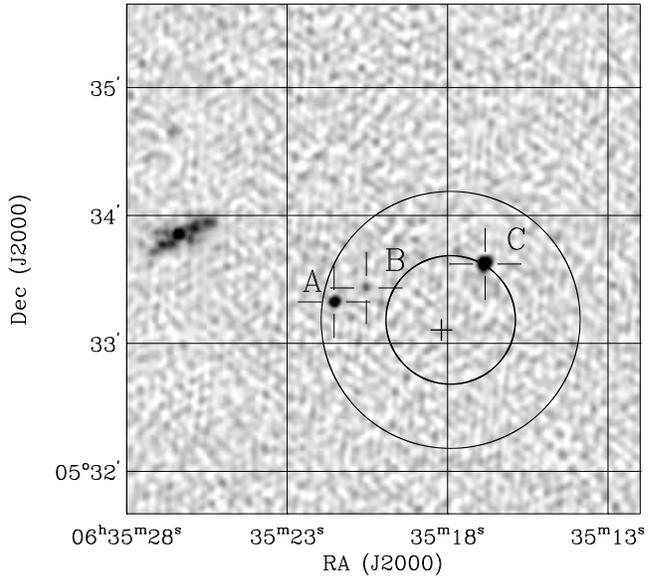}}
\caption[]{4.8 GHz VLA image of the region around \src. The $30''$ and
 $60''$ X-ray error circles (\cite{kaaret}; \cite{cusumano})
 and the position of the putative Be-star counterpart (+) are marked along
 with the 3 sources cited in the text.
}
\label{fig:6cm-img}
\end{figure}

If we adopt the 90\% confidence level $30''$ X-ray error radius circle, we
have only one source detected on its border while three sources are detected
within the $1'$ X-ray error circle quoted by Kaaret et al. (1999)
(see Fig. \ref{fig:6cm-img}).
The 4.8 GHz positions and fluxes of these sources are
given in Tab. \ref{tab:6cm-f}.
\begin{table}
\caption{4.8 GHz positions and fluxes of sources within the $60''$ X-ray
 error circle of \src.}
\label{tab:6cm-f}
\begin{tabular}{lccc}

Source  &  RA  &  Dec  & $F^*_{\rm 4.8\; GHz}$ \\
        &      &       & (mJy)  \\ \hline
 A & $06^{\rm h}\, 35^{\rm m}\, 21\fs64$ & $+05\deg\, 33'\, 19\farcs6$
 &  $0.79\pm 0.03$ \\
 B & $06^{\rm h}\, 35^{\rm m}\, 20\fs64$ & $+05\deg\, 33'\, 26\farcs4$
 &  $0.20\pm 0.03$ \\
 C & $06^{\rm h}\, 35^{\rm m}\, 16\fs94$ & $+05\deg\, 33'\, 37\farcs7$
  & $1.98\pm 0.04$ \\ \hline
\end{tabular}

*~Errors are $1\sigma$ level.~~Coordinates are equinox 2000.
\end{table}
At 1.4 GHz, only source C is detected with a flux of
$4.72\pm 0.15$ ($1\sigma$) mJy.
Assuming the flux spectrum can be described by a power law
$S\propto\nu^{\alpha}$, we find $\alpha=-0.72$ between 1.4 and 4.8 GHz.
We do not detect polarisation from any of these sources.

We checked for time variability of source C at 1.4 and 4.8 GHz
by accumulating 6 images for the 6 interleaved observations.
The source is constant within 25\% of its average flux.
We also produced the time series of 12 other field sources with 1.4 GHz fluxes
in the range 2--10 mJy.
They all show similar flux variations, probably due to calibration and
deconvolution uncertainties.
No optical counterpart to source C is visible in the POSS-II\footnote{
The Second Palomar Observatory Sky Survey (POSS-II) was made by the
California Institute of Technology with funds from the National Science
Foundation, the National Aeronautics and Space Administration,
the National Geographic Society, the Sloan Foundation, the Samuel
Oschin Foundation, and the Eastman Kodak Corporation.}
survey plates
at a limiting magnitude of $\sim 21$ (R) and $\sim 20$ (B).

We also carried out a search for radio pulsations at
430 MHz on 2000 January 20 using the Arecibo telescope.
The source was observed for 15 minutes with a 10-MHz bandwidth and a
sampling time of 0.1 $\mu$s using the AOFTM (Arecibo Observatory Fourier
Transform Machine).
The expected rms of these observations was 30 $\mu$Jy so that a $10\sigma$
detection corresponds to 0.3 mJy. The periodicity search in a wide period
range, covering the published 33.8 ms and its harmonics, did not result
in any convincing candidates above a $7\sigma$ threshold.
We did not detect any isolated dispersed pulses.

\section{Discussion}

\subsection{The nature of source C}
At 4.8 GHz we detect three sources in the $60''$ X-ray error circle quoted by
Kaaret et al. (1999); only source C is within the
reduced $30''$ error-circle by Cusumano et al. (2000) (Fig. \ref{fig:6cm-img}).
The steadiness of the flux of source A and its non-detection at 1.4 GHz,
suggesting an inverted spectrum, are not expected for an isolated neutron star.
While source B is too weak for any detailed analysis, we can constrain its
spectral index to be flatter than $-0.9$.
For source C, we estimate that the chance probability
to have a 1.4 GHz radio source of flux greater than 4.7 mJy
within a $30''$ error circle is 0.65\% ($2.7\sigma$) (\cite{windhorst}).
This value would suggest a putative
association but, on the other hand, it remains the only argument in favor.
In fact the lack of time variability, polarisation and pulsation and the flat
spectral index ($-0.72\pm0.05$ between 1.4 and 4.8 GHz), suggest an
extragalctic origin.

\subsection{The nature of the Be/X-ray pulsar}
We then accept the association of \src\ with the Be-star.
In this case the X-ray emission can be either accretion powered or
rotation powered (i.e. magnetospheric).
The observed X-ray properties (and the gamma-ray properties if one believes
the association with \gsrc) do not exclude this latter hypothesis.
We know that PSR B1259$-$63 (a Be-star/neutron star system with
a pulse period of 47 ms),
shows magnetospheric pulsed (away from periastron)
and continuum (near periastron) radio emission (see e.g. \cite{johnston96};
\cite{johnston99}), as well as variable unpulsed X-ray emission
(\cite{kaspi95}; \cite{nicastro}) along its $\sim 3.5$ years orbit.
This emission is thought to originate in a
discontinuity shock between the relativistic pulsar wind and the
Be-star wind (\cite{tavani}).
This same mechanism could account for the X-ray emission in \src.
The lack of radio flux could be due to scattering of the photons in a
high density Be-star wind, implying a short orbital period.
This hypothesis is supported by the high value of $N_{\rm H}$.

But if the radio emission is not quenched (at our flux limits)
by the local environment, then we must find another explanation for our
non-detections.
If we make the reasonable assumption we are dealing with a young 
($\tau \lapp 10^5$ years) isolated rotation-powered pulsar,
we then can estimate the energy loss rate, $\dot{E}$,
 from the X-ray conversion efficiency.
Extrapolating the flux in the 0.1--2.4 keV ROSAT band and assuming an
efficiency of 0.1\% (\cite{becker}), we obtain an
$\dot{E}\sim 1.5\times 10^{37}$ erg s$^{-1}$.
Frail \& Moffett (1993) found as average luminosity of young pulsars a
value $L_{\rm 1.4\; GHz}\simeq 30$ mJy kpc$^2$ and
$L_{\rm 400\; MHz}\simeq 300$ mJy kpc$^2$. They also calculate
a beaming fraction for such pulsars $f=61\% \pm 13\%$, significantly lower
than the $\sim 100\%$ value of X-ray emitting pulsars.
Assuming a distance to the system of 4 kpc, our observations would
imply a $5\sigma$ luminosity limit $L_{\rm 1.4\; GHz}\sim 10$ mJy kpc$^2$
and $L_{\rm 430\; MHz}\sim 5$ mJy kpc$^2$, then quite good.
For a distance of 4 kpc, we would expect a dispersion measure to this source
of $\sim 90$ pc cm$^{-3}$, given the Taylor \& Cordes (1993) model for
Galactic electron density. As the dispersion smearing across one of the
Arecibo receiver
10-kHz channels is only $\sim 0.1$ ms, our pulsed search was sensitive to the
shortest period pulsars.
Brazier \& Johnston (1999) argue that many young pulsars are visible as X-ray
pulsars but not as radio pulsars because they are
beaming away from Earth and not because they have intrinsically lower radio
fluxes.  Looking at their Fig. 1 we see
there are only a few radio pulsars or candidate neutron stars known with similar
$\dot{E}$ and luminosity, so it is likely that our non-detection could
also be attributed to an unfavorable beaming.

\subsection{Radio emission from accreting X-ray pulsars}

If \src\ is indeed an accretion-driven Be/X-ray binary pulsar,
as already noted by Cusumano et al. (2000), this would imply an unusually
low magnetic field strength $\lapp 10^9$ G similar to SAX J1808.4$-$3658
(\cite{gaensler}).
Given also the highly variable X-ray flux for the two systems,
it would be tempting to expect some transient non-magnetospheric radio
emission. Our $\sim 4$ hours observation time is much too short to say
anything in this respect.
A joint X-ray/radio monitoring program would be necessary to see if
any simultaneous, delayed or anti-correlated emission is present.
Fender et al. (1997b) showed
that there is a statistically significant anticorrelation between X-ray
pulsations and radio emission, though none of the 22 X-ray pulsars observed
in the centimeter band with a fairly good sensitivity
has ever been convincingly detected as a
synchrotron radio source (\cite{fender97a}).

Also, the lack of radio emission from the Be-star itself is not surprising
given its distance.
A survey of a sample of 13 optically and IR bright classical Be-stars
made from the Australia Telescope Compact Array failed to detect any star at
4.8 GHz (\cite{clark}) with a $3\sigma$ flux level of
$\sim 0.1$ mJy.

\section{Conclusions}
We have performed sensitive radio observation of \src\ at 1.4 and 4.8 GHz
with the VLA and pulsar search at 430 MHz with the Arecibo telescope.
We do not detect any flux at the Be-star position, but a relatively
strong, flat spectrum source lies within the $30''$ X-ray error circle.
Its properties do not support the association with the X-ray
source and no sign of pulsed emission is found at a relatively high
sensitivity limit.
Then, assuming the system is a Be/X-ray pulsar, the lack of radio emission
at our flux limits does not allow us to distinguish between an
accretion or rotation-powered system. However it shows that:
\begin{itemize}
\item the system probably has a short binary period and
 therefore it is quite different from PSR B1259$-$69;
\item if periodic/flaring radio emission 
 is present, our observation was far too short to be
 conclusive in this respect;
\item the radio emission could be quenched by the high density Be-star wind
 or the pulsar is beaming away from the Earth;
\item the system could be an accreting X-ray pulsar, though transient
 non-magnetospheric weak radio emission could still be present.
\end{itemize}
 
Given the variable nature of the X-ray source it is also unlikely that any
accurate timing can be performed in order to check for gamma-ray emission
at 33.8 ms from \gsrc.
However its gamma-ray spectral index and flux stability
favor an isolated neutron star.
Until the arrival of more sensitive
gamma-ray instruments such as GLAST, the association between the two sources
will likely remain uncertain.
As Kaaret et al. (1999) and Cusumano et al. (2000),
we conclude that \src\ is most likely a compact
Be-star/neutron-star X-ray binary pulsar, though with an unusually short pulse
period.
Further X-ray observations are required both to confirm the 33.8 ms period
(and measure a spin-up/down) and to determine the (likely short) binary period.

After this paper was submitted we became aware of new X-ray results by
Kaaret et al. (2000) which suggest an orbital period of $\simeq 11$ days
and a large
$\dot{\nu}$ which is more typical of young, rotation-powered pulsars
rather than of accretion-powered X-ray binaries.

\begin{acknowledgements}
The National Radio Astronomy Observatory is a facility of the National Science
Foundation.
L.N. acknowledges the support of CNR through a short-term mobility program.
B.M.G. acknowledges the support of NASA through Hubble Fellowship
grant HST-HF-01107.01-A awarded by the Space Telescope Science Institute
which is operated by the Association of Universities for
Research in Astronomy, Inc., for NASA under contract NAS 5-26555.

\end{acknowledgements}

\end{document}